\documentstyle[natbib,epsfig]{elsart}

\begin{document}

\runauthor{D.A.Harris, A.Para}


\begin{frontmatter}


\title{Neutrino Oscillation 
Appearance Experiment using Nuclear Emulsion and Magnetized Iron}




 \author[inst1,inst2]{D.A.Harris} 
 \author[inst2]{, A. Para} 
 \address[inst1]{University of Rochester, Rochester, NY 14853} 
 \address[inst2]{Fermi National Accelerator Lab, Batavia, IL 60510} 



\begin{abstract}
This report describes an apparatus that could be used to 
measure both the identity and charge of an outgoing lepton 
in a charged current neutrino interaction.  This capability 
in a massive detector would allow  the most comprehensive 
set of neutrino oscillation physics measurements.  By measuring 
the six observable transitions between initial and final state
neutrinos, one would be able to measure all elements 
of the  neutrino mixing matrix, as well as search for 
CP violation, and matter effects.  If the measured matrix is not 
unitary, then one would also have an unambiguous determination of 
sterile neutrinos.  Emulsion is 
considered as the tracking medium, and different 
techniques are discussed for the application of a 
magnetic field.  
\end{abstract}




\end{frontmatter}






\section{Introduction} 

To observe all of the possible 
neutrino oscillation phenomena one would ideally ask for pure neutrino beams
of different flavors, and a detector
which could identify the three possible leptons in the final state 
of a charged current interaction. Traditional neutrino beams contain predominantly
 $\nu_{\mu}$ or   $\overline\nu_{\mu}$ with a small admixture of other neutrino
flavors.
 Muon storage rings offer a possibility
of mixed $\overline\nu_{\mu}$/$\nu_{e}$ and $\nu_{\mu}$/$\overline\nu_{e}$
beams. These beams will have virtually no admixture of other neutrino flavors,
but they will contain neutrinos and antineutrinos.
 To exploit fully these beams and uniquely identify the oscillation 
mode one needs a determination of the electric charge of the outgoing lepton.
This will uniquely distinguish  neutrino from antineutrino interactions. 
    
Several massive detectors exist which can identify the presence and 
charge of an outgoing muon, and some have been proposed to detect 
the presence of an outgoing tau or electron.  However, measuring both the 
presence AND the charge of an outgoing tau, or electron, on an event-by-event basis
 remains 
a serious challenge in the field of neutrino oscillation experiments.
Such a measurement is necessary to detect a 
$\overline\nu_{\mu}\rightarrow\overline\nu_{e}$ oscillation in the presence of a
$\nu_{e}$ component of the beam and/or to distinguish
$\nu_{\mu}\rightarrow\nu_{\tau}$ from 
$\overline\nu_{e}\rightarrow\overline\nu_{\tau}$ oscillations.
Precise measurements of $\nu_{\mu}\rightarrow\nu_{\tau}$, 
$\nu_{\mu}\rightarrow\nu_{e}$ and $\nu_{e}\rightarrow\nu_{\tau}$ 
oscillation amplitudes would allow a test of unitarity 
 of the neutrino oscillation mixing matrix, and an indirect search 
for sterile neutrinos. A measurement of the difference between 
$\nu_{\mu}\rightarrow\nu_{e}$ and
$\nu_{e}\rightarrow\nu_{\mu}$ oscillation amplitudes could provide 
a direct test of T (or CP) violation, free of matter effects which 
affect a  $\nu_{e}\rightarrow\nu_{\mu}$ and 
$\overline\nu_{e}\rightarrow\overline\nu_{\mu}$ comparison.

In this paper we describe a detector which combines
emulsion technology with a magnetic field. High resolution tracking capabilities
of nuclear emulsions permit unambiguous detection of produced $\tau$ leptons and
electrons, whereas a measurement of the deflection in a magnetic field allows 
a determination of the sign of the electric charge.

\section{ Lepton Identification in the Emulsion Detector } 

The CHORUS experiment has  demonstrated the capabilities for identification
of $\tau$ leptons through observation of its decay kinks in emulsion, and has set 
the most stringent limits on $\nu_\mu \to \nu_\tau$  at high $\delta m^2$ 
\cite{ref1} to date.  CHORUS uses a bulk nuclear emulsion as a detector; 
such a  technique is prohibitively expensive for very large mass detectors.


A different geometry which is being studied by both the MINOS and
OPERA collaborations involves interspacing  emulsion plates, used
as detectors, with thin lead plates, used as a target\cite{edgecock}\cite{ref2a}.
This geometry was used succesfully in studies of cosmic rays by the JACEE 
collaboration\cite{JACEE}.
Electronic tracking devices, sampling the emulsion detector with frequency
of the order of $0.5\lambda_{I}$ are used to trigger the detector and to
locate the interaction point to a small volume.   That small volume can be removed
from the detector and analyzed in nearly real time, with electronic tracking information
serving as a  guide to the analysis of the emulsion sheets.

An emulsion  detector optimized for tau detection typically involves a thin lead
plate serving as a target, followed by a a gap with two emulsion layers 
separated by a low-Z-material spacer. Track segments and their spatial angles 
are measured in both emulsion  layers. Tau decays inside the spacer 
material are characterized by a large angle (typically above $50 mrad$) between the
downstream and upstream track segments. A certain fraction of taus decaying in the
target plate can be identified by a large impact parameter of the resulting tau 
daughter.


Electromagnetic showers are sampled in emulsion with
a typical granularity of the order of  $0.2 X_0$. Thanks to the 
excellent spatial resolution
of the emulsion and a high sampling frequency of the electromagnetic shower,
individual acts of photon emission and conversion are easily detectable in
the emulsion. Additional information is provided by the double ionization 
(measured via grain density along the track) of the electron-positron pair from
a photon conversion.

An electron can be thus identified as a charged track with several conversion pairs 
within a small cone around it. An important feature of the emulsion detector is its 
ability to follow the initial electron track even inside the electromagnetic cascade.

\section{Lepton Charge Determination}

The electric charge of the particles can be determined by measuring the trajectory of
a particles in the magnetic field. The simplest solution consists of replacing 
 the lead target plates by iron ones and using an external coil to create a magnetic
field in the iron in excess of $1 Tesla$.

Charged particles traversing  the iron plate will receive a  $p_t$ kick
of $0.003 x B GeV$, where  $x$ is the thickness of the steel plate in $cm$ and
$B$ is the magnetic field in $Tesla$.
At the same time, however, multiple scattering in the iron will generate 
a random pt of 
$0.014GeV \sqrt{\frac{x}{X_0}}$, where $X_0=1.76cm$ is the radiation length
of steel. For a typical field strength of $1 Tesla$, the multiple scattering 
effects dominate over the bending in the magnetic field. The situation  improves when 
several iron plates are traversed: the $p_t$ due to the magnetic field add linearly, whereas 
the  multiple scattering induced $p_t$ grows like $\sqrt{x}$. After a traversal
of $N$ steel plates with thickness $x$ each, the significance of the charge 
determination is 
\begin{equation} 
\sigma = 2\frac{ 0.003 B(Tesla) x(cm) }{0.014 \sqrt{\frac{x}{X_0}}} 
\sqrt{N} 
\end{equation}
The factor of $2$ accounts for the fact that we are not trying to determine the 
sign as such, but rather to distinguish between a positive and negative track hypothesis.

For a typical example of a detector with $x=1 mm$ ($0.05X_0$) of steel, 
in a $1 Tesla$ field, it takes about $100$ planes or $10 cm$ of iron to achieve
a 2 $\sigma$  sigma sign measurement. 
For muons, or even daughters of tau leptons it should be possible
to achieve $3$ or $4 \sigma$ sign determination. In the electron case it may
become impractical to follow the primary electron beyond some $5-10 X_0$, however.

In case a more precise charge determination is required for electrons, the solution
could be to immerse the entire detector in an external, strong magnetic field.
In such a case a significant improvement of the measurement is achieved, as the
space between the steel plates contributes to the magnetic bending as well, whereas
its contribution to the multiple scattering is negligible in comparison with that of
the iron plates.

As an example,  consider a detector with $x=1 mm$ steel plates and a 
$2 mm$ gap between them, immersed in $4 Tesla$ magnetic field: it will require
only $4-5$ emulsion plates to achieve a $5 \sigma$ charge measurement. 

Such a detector is not an impractical one: a 1 kton  detector would have a total
volume of $375 m^{3}$ ($5 m \times 5 m \times 15 m$). This is a rather modest volume in 
comparison, for example, with the ATLAS barrel toroidal magnet system\cite{atlas}, 
which generates a magnetic field of $2.4 - 4.2 Tesla$ in a total volume of $7600 m^{3}$. The ultimate size of a possible detector will be, thus, limited by the available 
resources, rather than by technical factors. A cost of a  hypothetical detector with
 20 kton of active mass would be of the order of 1100 million SF: 100 MSF for
the superconducting magnet and 1000 MSF for the emulsion detector. 

\section{ Conclusion } 

We have presented a concept of a novel detector involving
large volume emulsion detectors interspaced with thin layers of  steel.
Such a detector should be capable of identification of the final state
lepton (muon, electron or tau) with good efficiency and very small background.
The superb spatial resolution and  granularity of nuclear emulsions makes it possible to
determine the sign of the electric charge of the lepton, once the detector is
immersed in the magnetic field. The magnetic field can be generated by the
excitation of  steel plates with an external coil, or by immersion of the
entire detector inside a superconducting magnet, similar to the ones being
built for the LHC experiments.

With such a detector one can take full advantage of the physics opportunities
provided by intense neutrino beams produced at a muon storage ring. It will allow
a precise determination of the neutrino mixing matrix elements, search for matter and
CP violation effects.
A very large mass detector, of the order of 20 kton, can be constructed at the price
of the order of 1 BSF. While this cost is probabbly comparable with the cost of
construction of the neutrino factory itself, this detector would provide as complete
a set of physics information as  possible.





\begin{thebibliography}{999}

\bibitem{ref1} E.Eskut {\em et al.}, Nucl.Inst. and Meth.A401,7(1997);	
see also: CHORUS collaboration, submitted to 
Lepton Photon 99, July 1999, hep-ex/9907015.  

\bibitem{edgecock} MINOS Collaboration, Fermilab proposal P-915, NuMI-E-473, April 1999; see also R. Edgecock, these proceedings.  

\bibitem{ref2a} Migliozzi, these proceedings.

\bibitem{JACEE} Burnett, T. H. {\em et al.}, NIM, {\bf A251} , p.583, (1986). 
\bibitem{atlas} W.W. Armstrong {\em et al}, Atlas Technical Proposal.  
\end{thebibliography}
\end{document}